# Quantum Emitters in Aluminum Nitride Induced by Zirconium Ion Implantation

Alexander Senichev[1,2], Zachariah O. Martin[1,2], Yongqiang Wang[2], Owen M. Matthiessen[1], Alexei Lagutchev[1], Han Htoon[2,4], Alexandra Boltasseva[1,2], Vladimir M. Shalaev[1,2*]

[1]Elmore Family School of Electrical and Computer Engineering, Birck Nanotechnology Center and Purdue Quantum Science and Engineering Institute, Purdue University, 610 Purdue Mall, West Lafayette, IN, 47907, USA
[2]Quantum Science Center, Department of Energy, A National Quantum Information Science Research Center of the U.S, Oak Ridge National Laboratory, 1 Bethel Valley Road, Oak Ridge, TN 37830, USA
[3]Materials Science and Technology Division, Los Alamos National Laboratory, Los Alamos, NM, 87545, USA
[4]Materials Physics and Applications Division, Los Alamos National Laboratory, Los Alamos, NM, 87545, USA
*Email: shalaev@purdue.edu

Keywords: quantum photonics, single-photon emitters, aluminum nitride, integrated photonics, zirconium, krypton, ion implantation

The integration of solid-state single-photon sources with foundry-compatible photonic platforms is crucial for practical and scalable quantum photonic applications. This study investigates aluminum nitride (AlN) as a material with properties highly suitable for integrated on-chip photonics specifically due to AlN capacity to host defect-center related single-photon emitters. We conduct a comprehensive study of the creation and photophysical properties of single-photon emitters in AlN utilizing Zirconium (Zr) and Krypton (Kr) heavy ion implantation and thermal annealing techniques. Guided by theoretical predictions, we assess the potential of Zr ions to create optically addressable spin-defects and employ Kr ions as an alternative approach that targets lattice defects without inducing chemical doping effects. With the 532 nm excitation wavelength, we found that single-photon emitters induced by ion implantation are primarily associated with vacancy-type defects in the AlN lattice for both Zr and Kr ions. The emitter density increases with the ion fluence, and there is an optimal value for the high density of emitters with low AlN background fluorescence. Additionally, under shorter excitation wavelength of 405 nm, Zr-implanted AlN exhibits isolated point-like emitters, which can be related to Zr-based defect complexes. This study provides important insights into the formation and properties of single-photon emitters in aluminum nitride induced by heavy ion implantation, contributing to the advancement of the aluminum nitride platform for on-chip quantum photonic applications.

## Introduction

Photonic platforms hold a great potential for the realization of practical quantum technologies spanning communication, computing, and sensing [1] [2] [3] [4] [5]. Photons serve as excellent qubits for quantum networks, long-distance entanglement distribution, and quantum information processing due to their numerous degrees of freedom, high propagation speed, and resistance to decoherence [6] [7] [8]. Photonic integrated circuits (PICs) with compact, controllable, and reconfigurable elements are crucial for development of complex quantum systems [9] [10] [11]. On-chip incorporation of quantum light sources like single-photon emitters is a key to this integration [12] [13]. Solid-state single-photon emitters offer reliable, potentially deterministic sources of single photons with high repetition rates and indistinguishability, enabling quantum logical operations and entanglement distribution [14]. Importantly, quantum emitters in solids can provide the necessary interface between spin qubits and photons [15].

Combining materials with single-photon emitters such as III-V quantum dots [16] [17] [18], diamond [19] [20], hexagonal boron nitride [21] [22] [23], and photonic platforms including silicon, silicon nitride, aluminum nitride, or lithium niobate [24] [25] [26], is currently accomplished predominantly by hybrid integration. However, there is a growing interest in alternative material platforms that can host both intrinsic high-purity single-photon sources and are compatible with wafer-scale growth and foundry fabrication processes. Such photonic platforms with intrinsic quantum emitters include silicon [27] [28] [29], silicon carbide [30] [31] [32], silicon nitride [33] [34] [35], aluminum nitride [36] [37] [38], and gallium nitride [39] [40] [41], to name a few.

Among these platforms, aluminum nitride stands out for its unique properties, including compatibility with metal–oxide–semiconductor (CMOS) fabrication processes, wide-bandgap (6.2 eV), high refractive index (2.1), wide transparency window (from ultraviolet to mid-infrared), as well as second-order nonlinearity ($\chi^{(2)}$ = 4.7 pm/V) [42], though weaker compared to lithium niobate ($\chi^{(2)}$ = 41.7 pm/V) [43]. AlN-based waveguides also exhibit moderately low losses of about 0.4 dB/cm around 1550 nm [44]. Even lower losses of ~ 1 dB/m were achieved in $Si_3N_4$-based waveguides, [45], however they suffer from strong autofluorescence in the visible spectral range, essentially absent in AlN [46]. Such autofluorescence can hinder the detection of single-photon emission from defects. The non-centrosymmetric wurtzite crystal structure of AlN possesses a piezoelectric property, which makes it suitable for microelectromechanical systems (MEMS) [47]. AlN-based piezo-optomechanical actuators are demonstrated for cryogenic compatible programmable photonic circuits with high-speed phase modulation [48]. For more details on the properties of AlN for PICs, reviews [46] [47] [49] contain a comprehensive list of recent advances in on-chip AlN photonic devices. Finally, III-Nitride semiconductors, including AlN, play a crucial role in solid-state lighting technology. They are second only to silicon within the microelectronics industry, benefiting from well-established growth processes and device fabrication techniques and facilities.

AlN has also attracted a particular attention as a host of intrinsic point defects that act as single-photon sources [36] [37] [38] [50]. Single-photon emitters in AlN were observed in as-grown samples fabricated by metal organic chemical vapor deposition [36]. The formation of quantum emitters has also been observed in AlN films after implantation with helium (He) ions and thermal annealing [38]. Recently, it was shown that the femtosecond laser writing can be applied to create single quantum emitters in AlN crystal with a creation yield of > 50% [51].

Despite the numerous single-photon emitters identified in AlN, none have yet demonstrated to possess individually addressable spin properties. First-principles calculations suggest that AlN can indeed host spin defects, with Titanium (Ti) and Zirconium (Zr) transition-metal impurities [52] [53]. These are predicted to form stable complexes with nitrogen vacancies, resulting in defects with spin-triplet ground states [52] [53]. Incorporation of Zr species into AlN by ion implantation has been recently studied experimentally [54]. It was shown that Zr implantation into AlN induces an additional emission peak at 730 nm, attributed to Zr-based $(Zr_{Al}–V_N)^0$ defects. The previous structural analysis of Zr-implanted epitaxial AlN films by Raman and X-ray absorption spectroscopies indicated that the dominant defects were a substitutional impurity $Zr_{Al}$ or vacancies such as $V_{Al}$ and $V_N$ [55]. However, the single-photon properties of these defects remain unexplored.

In this study, we generated single-photon emitters in AlN through the implantation of heavy ions followed by thermal annealing. Zr ions were chosen to experimentally test the theoretically predicted creation of Zr-based quantum emitters with optically addressable spin properties. As a control, Kr ions were employed to create quantum emitters based on crystal-lattice defects without inducing chemical doping for comparison with Zr-implanted samples. A broad ion beam implantation was applied to simplify the search of created emitters and assess their density. We systematically varied ion implantation fluences to obtain the desired density of emitters without a drastic damage of the crystal lattice.

Our results show that, under 532 nm excitation wavelength, the observed single-photon emitters are primarily associated with vacancy- or interstitial-type defects in the AlN lattice induced by ion bombardment, for both Zr and Kr ions, without their incorporation. We observed a broad spectral distribution of emission from individual single-photon sources with the appearance of narrow emission peaks at room temperature, as compared to emitters created in He-implanted AlN [38]. Additionally, under shorter wavelength excitation at 405 nm, Zr-implanted films exhibited isolated point-like defects emitting in the spectral range suggested for Zr-based spin defects. This work lay the ground for further experiments to verify the correlation of these single-photon emitters with Zr-based defect complexes and address their spin properties.

**Experimental**

For this study, we used commercially available crystalline 200-nm-thick AlN films grown on sapphire (refer to Methods for detailed information). The fabrication process of quantum emitters by heavy ion implantation is illustrated in **Fig. 1a**. Prior to ion implantation and thermal annealing the 2-inch AlN-on-Sapphire wafers were diced into 10x10 mm$^2$ samples. Photoluminescence intensity maps of as-grown AlN samples recorded before ion implantation show low background fluorescence and the absence of isolated emitters (**Fig. 1d**). Two groups of samples were implanted with $^{90}Zr^+$ ions at implantation energy of 200 keV and $^{84}Kr^+$ ions at 190 keV, respectively. The implantation energy was chosen using Stopping and Range of Ions in Matter (SRIM) simulations to obtain the implantation depth of approximately 100 nm or less (**Fig. 1b**). The implantation energies were kept the same for all the implanted samples. The implantation fluence was systematically varied to assess its influence on the formation of single-photon emitters. We selected four ion fluences detailed in **Table 1**, which cover the available range of the 200 kV Danfysik Research Ion Implanter (see Methods for details).

The as-grown sample A, Zr-implanted samples B-E, and Kr-implanted samples F-I were annealed under the same conditions: at 1000°C for 30 minutes in an argon atmosphere (**Fig. 1c**). The thermal annealing step was performed to repair the lattice damages introduced by ion implantation and to activate single-photon emitters (**Fig. 1e**).

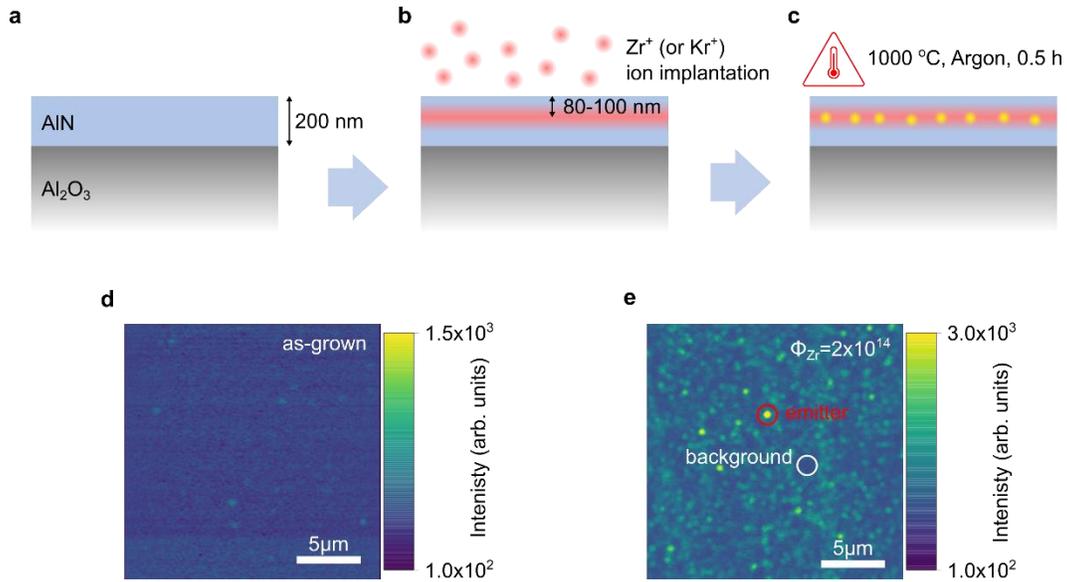

**Figure 1**. Fabrication process of single-photon emitters in AlN by heavy ion implantation using Zr and Kr: (a) As-grown 200-nm-thick AlN film on sapphire grown by a plasma vapor deposition nanocolumnar process (Kyma Technologies Inc.); (b) Zr and Kr ion implantation with an acceleration voltage of 200 keV and 190 keV, respectively, for the target implantation depth of about 80-100 nm; (c) Annealing of ion-implanted AlN films at 1000°C in argon ambience for 30 minutes for creation of photostable single-photon emitters. (d) PL intensity map of an as-grown AlN sample without thermal annealing. (e) PL intensity map of a representative AlN film implanted at a Zr-ion fluence of $\Phi_{Zr} = 2 \times 10^{14}$ (sample D) and annealed at 1000°C in argon atmosphere for 30 minutes. Red circle: representative single-photon emitter, white circle: typical area with no emitters for AlN background emission measurements.

**Table 1**. List of samples with applicable implantation and thermal annealing conditions

| Sample # | Implantation Species | Implantation Energy | Ion fluence $\Phi_{Zr}$, ions/cm² | Thermal annealing |
|---|---|---|---|---|
| A | Control sample without implantation | | | |
| B | $^{90}$Zr$^+$ | 200 keV | $5 \times 10^{11}$ | T = 1000°C Argon 30 min |
| C | | | $1 \times 10^{13}$ | |
| D | | | $2 \times 10^{14}$ | |
| E | | | $2 \times 10^{15}$ | |
| F | $^{84}$Kr$^+$ | 190 keV | $6 \times 10^{11}$ | |
| G | | | $1 \times 10^{13}$ | |
| H | | | $2 \times 10^{14}$ | |
| I | | | $2 \times 10^{15}$ | |

## Results and discussion

### Emergence of single-photon emitters in ion-implanted AlN films

First, we examine the density of single-photon emitters in AlN following thermal annealing as a function of the ion implantation fluence. We focus on the results from Zr-implanted AlN samples, discussed in the main manuscript. Supplementary Materials detail the results from Kr-implanted samples, providing a reference for a comparative analysis.

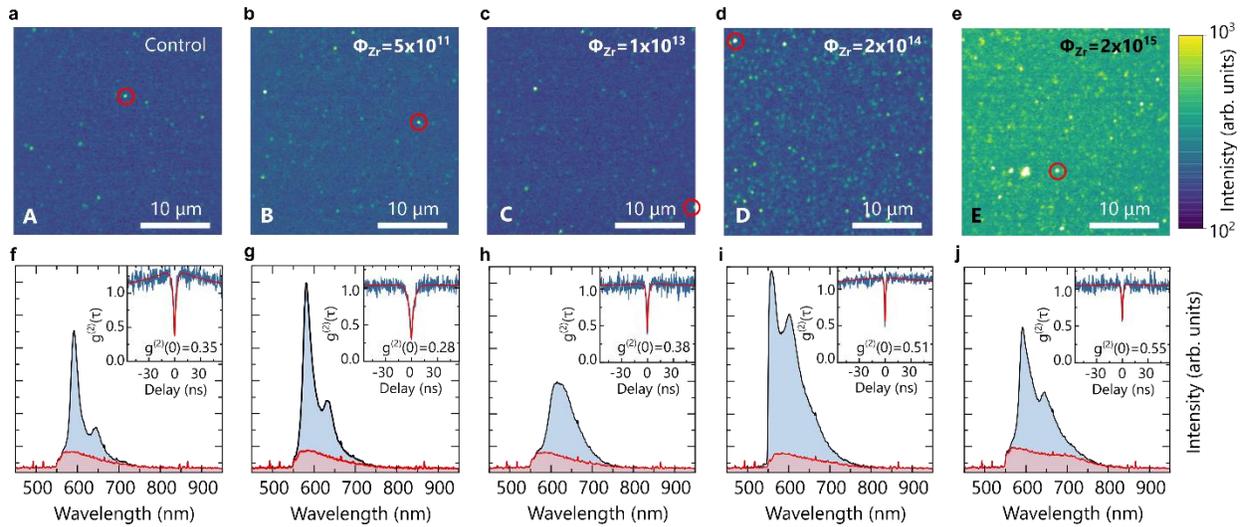

**Figure 2**. Confocal 30x30 µm² PL intensity maps for different Zr-implanted AlN samples with an implantation fluence: (a) as-grown control sample, (b) $\Phi_{Zr} = 5 \times 10^{11}$, (c) $\Phi_{Zr} = 1 \times 10^{13}$, (d) $\Phi_{Zr} = 2 \times 10^{14}$, (e) $\Phi_{Zr} = 2 \times 10^{15}$. (f-j, in blue): PL spectra of representative emitters indicated in PL intensity maps (a-e) with red circles. (f-j, in red): background PL spectra of AlN film in the areas without emitters. Insets: second-order autocorrelation histograms $g^{(2)}(\tau)$ fitted with a three-level model.

In search for single-photon emitters in AlN films, we employed a 532 nm laser, commonly used for studying quantum emitters in the visible range. We found that subsequent annealing of the as-grown control sample A without implantation resulted in sparse isolated bright spots evidenced in PL intensity maps (**Fig. 2a**). These were confirmed to be single-photon emitters by second-order autocorrelation measurements. Our observations agree with earlier findings in the literature, suggesting that as-grown AlN films may contain intrinsic point defects capable of acting as single-photon emitters [36] [37]. The PL spectrum and the second-order autocorrelation histogram of a representative quantum emitter from the as-grown annealed AlN sample (A) are shown in **Fig. 2f**. A typical PL spectrum from one of these emitters consists of a pronounced emission peak and a red-shifted broad emission attributed to the zero-phonon line (ZPL) and phonon sideband (PSB), respectively. The PL intensity of emitters is substantially stronger than background PL as can be seen from the comparison of two corresponding spectra (**Fig. 2f** blue and red curves, respectively). Fitting of the second-order autocorrelation histogram yields a $g^{(2)}(0)$ value of 0.35 without background correction or spectral filtering.

Next, we analyze the properties of single-photon emitters created by ion implantation. We found that the density of emitters changes as a function of the ion implantation fluence (**Fig. 2b-e**). The density of the isolated emitters remains virtually the same for two lowest ion fluences of $5 \times 10^{11}$ ions/cm² (sample B) and $1 \times 10^{13}$ ions/cm² (sample C) and is comparable to the as-grown control sample (A) after annealing (**Fig. 2a-c**). However, the emitter density increases by at least a factor of two for the implantation fluence of $2 \times 10^{14}$ ions/cm² (sample D). Further increase of the implantation fluence to $2 \times 10^{15}$ ions/cm² (sample E) do not substantially change the density of emitters, but dramatically increase the background fluorescence of the AlN film itself. Hence, the ion fluence of $2 \times 10^{14}$ ions/cm² or near its vicinity appeared to be optimal for creating emitters with high surface density while simultaneously keeping the AlN background fluorescence close to the level of as-grown AlN.

PL spectra of representative single-photon emitters in AlN for four implantation fluences are shown in **Fig. 2g-j**. The PL spectrum structure of individual emitters resembles the one observed in the control sample A with a pronounced ZPL peak accompanied by lower energy sidebands, PSBs. The room-temperature ZPL linewidth varies among different emitters, which may be related to different implantation conditions, but establishing a correlation will require further analysis. The typical $g^{(2)}(\tau)$ values at zero delay time $\tau = 0$ registered without spectral filtering or background correction were typically below or about 0.5, confirming the single-photon nature of the emitters (insets of **Fig. 2g-j**). The decrease of the single-photon purity with the increase of the implantation fluence to $2 \times 10^{15}$ ions/cm² may be attributed to the increased background fluorescence of the AlN films (**Fig. 2j**, inset).

The investigation into Kr-implanted AlN films reveals a comparable correlation between the formation of single-photon emitters and implantation fluence, resulting in increased emitter density and elevated background fluorescence as the fluence increases (**Fig. S1**, Supplementary Materials). Notably, the background intensity seems lower for Kr-implanted AlN films in comparison to the Zr-implanted samples, which we will discuss later.

Along with typical PL emission spectra presented in **Fig. 2g-j**, occasionally PL spectra with different features, like a single broad peak or multiple peaks of varying widths and relative intensities were also observed (**Fig. S2**, Supplementary Materials). These less common spectra may result from the interaction of quantum emitters with the surrounding AlN lattice, the presence of multiple emitters within the excitation spot, or the existence of additional defect complexes associated with various point-like defects, including vacancies, substitutions, or impurities. Notably, these spectra differ from previously reported ones for quantum emitters in as-grown AlN with relatively weak ZPL transition and an intense broad phonon sideband [37]. Furthermore, certain AlN defects created by our method produce emission peaks with narrower linewidths as compared to those reported for He-implanted AlN defects at room temperature [38].

**Spectral properties of the photoluminescence from AlN single-photon emitters**

We now turn to a comparative analysis, aiming to differentiate between pre-existing defects in as-grown sample A and those induced by ion implantation in both Zr-implanted (B-E) and subsequently Kr-implanted AlN (F-I). For ion-implanted samples, we focus on the sample with the optimal implantation fluence $\Phi_{Zr} = 2 \times 10^{14}$ (D), which exhibited the highest emitter density and minimal AlN background fluorescence.

We analyzed the distribution of ZPL wavelength for the control and ion implanted samples (**Fig. 3a, d**). For this purpose, we identified single-photon emitters with the $g^{(2)}(0) < 0.5$ (**Fig. 3c, f**) and recorded the wavelength of the most intense peak per spectrum, attributed to the ZPL transition. This analysis shows similar median values of the ZPL peak position: 592 nm and 590 nm for the control (A) and implanted sample (D), respectively. Both histograms reveal the most frequent value for ZPL at 585 nm. Interestingly, the ZPL histogram peak at this spectral position was also reported for MOCVD-grown AlN films [36]. In addition, a possibly less pronounced peak at roughly 630 nm was resolved in our histograms.

Representative PL spectra of emitters from the spectral regions indicated with dashed lines in **Fig. 3a, d** for as-grown and Zr-implanted samples are shown in **Fig. 3b, e**. The room-temperature ZPL linewidth, assessed using the Lorentzian line shape, ranges from 8 nm to 17 nm. Sample D implanted with $\Phi_{Zr} = 2 \times 10^{14}$ also shows similar PL spectra with ZPL linewidths spanning 9 nm to 24 nm, not significantly wider than emitters in the as-grown AlN.

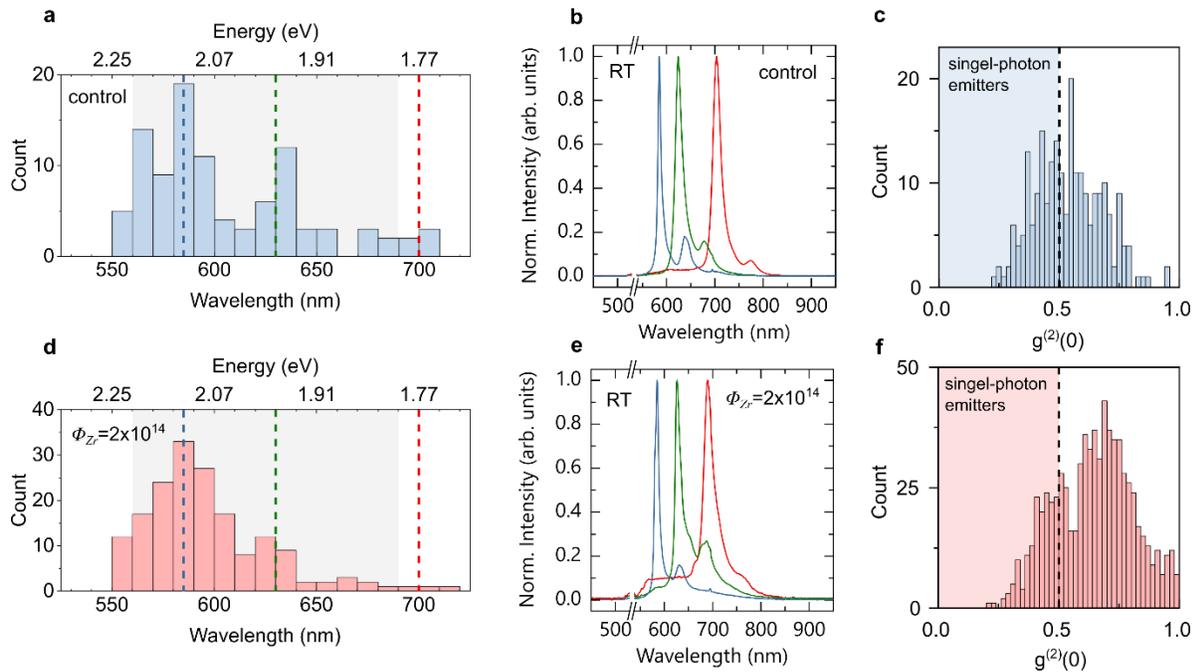

**Figure 3**. (a, d) ZPL peak wavelength distribution obtained from 102 and 174 single-photon emitters with the $g^{(2)}(0)<0.5$ in the control (A) and implanted (D) samples, respectively; bin size 10 nm. (b, e) Representative PL spectra for spectral regions indicated in (a, d). (c, f) Histograms of $g^{(2)}(0)$ distribution for emitters in the sample A (238 emitters) and sample D (745 emitters); bin size 0.02. The $g^{(2)}(0)$ values were obtained without background correction and spectral filtering.

The wavelength distribution of the most intense peak in PL spectra of Kr-implanted is shown in **Fig. S3**, Supplementary materials, and follows similar behavior as for control and Zr-implanted samples. This points to possibility of defects created in both samples to be of similar nature. It appears that both Zr and Kr ions were likely to induce damage to the AlN lattice and form vacancy-type defects, rather than becoming chemically bonded to it.

In addition to room-temperature measurements, we performed PL spectra measurements of Zr-implanted samples at cryogenic temperatures. At liquid-nitrogen temperature the single-photon emitters exhibited substantially narrower ZPL linewidths - down to about 4 nm. Moreover, the emission was found to predominantly occur through the ZPL transitions. We estimated the Debye-Waller factor to be about 22, 43, and 62% for PL spectra of SPE1, SPE2, and SPE3 emitters shown in **Fig. 4a**, respectively.

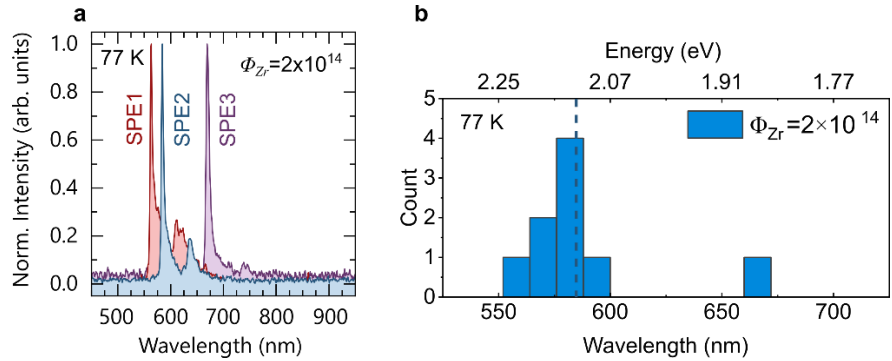

**Figure 4**. Low-temperature (77 K) photoluminescence spectroscopy of single-photon emitters in AlN. (a) Three representative spectra with ZPL wavelength of $SPE_1$ 563 nm, $SPE_2$ 584 nm, and $SPE_3$ 669 nm. (b) ZPL wavelength distribution for nine PL spectra measured at 77K with the mean value of 588±32 nm, dashed line; bin size 12 nm.

Analyzing the single-photon purity of AlN quantum emitters at room temperature (**Fig. 3c, f**), we found that the $g^{(2)}(0)$ histogram from the as-grown sample A shows a distribution with a median value of 0.54. In Zr-implanted samples, a substantial number of emitters exhibited $g^{(2)}(0)$ values below 0.5, although the median value of the respective distribution is 0.65. A similar trend is observed for emitters in Kr-implanted samples (**Fig. S3**, Supplementary Materials). This can be attributed to the presence of ion-induced emitters in the implanted samples that differ from those in the as-grown samples. The reduced $g^{(2)}(0)$ values are due to either the defect structure itself or an increased background emission rate.

**Polarization analysis of emission**

We measured PL intensity maps from the Zr-implanted sample D for various orientations of the polarizer in front of the detector, e.g., **Fig. 5a** at 0°. The map of the difference between PL intensities measured at 0° and 90° allowed us to identify emitters with different emission polarizations (**Fig. 5b**). Additional maps of the difference between PL intensities measured at different orientations of the polarizer are shown in Supplementary Materials, **Fig. S4**. The polarization diagrams in **Fig. 5c-d** show quantum emitters with mutually orthogonal polarization states, consistent with previously reported emitters in He-implanted AlN samples [38]. Moreover, we observed emitters with varied PL polarization (**Fig. 5e**), suggesting the presence of various emitter orientations. The PL spectra of these emitters are shown in **Fig. 5f-h**. Further investigating the correlation between crystalline structure orientation and polarization of the photoluminescence requires in-depth analysis beyond the scope of this study. Such future investigation could uncover emitter classes and corresponding defects correlated with crystallographic directions in AlN films.

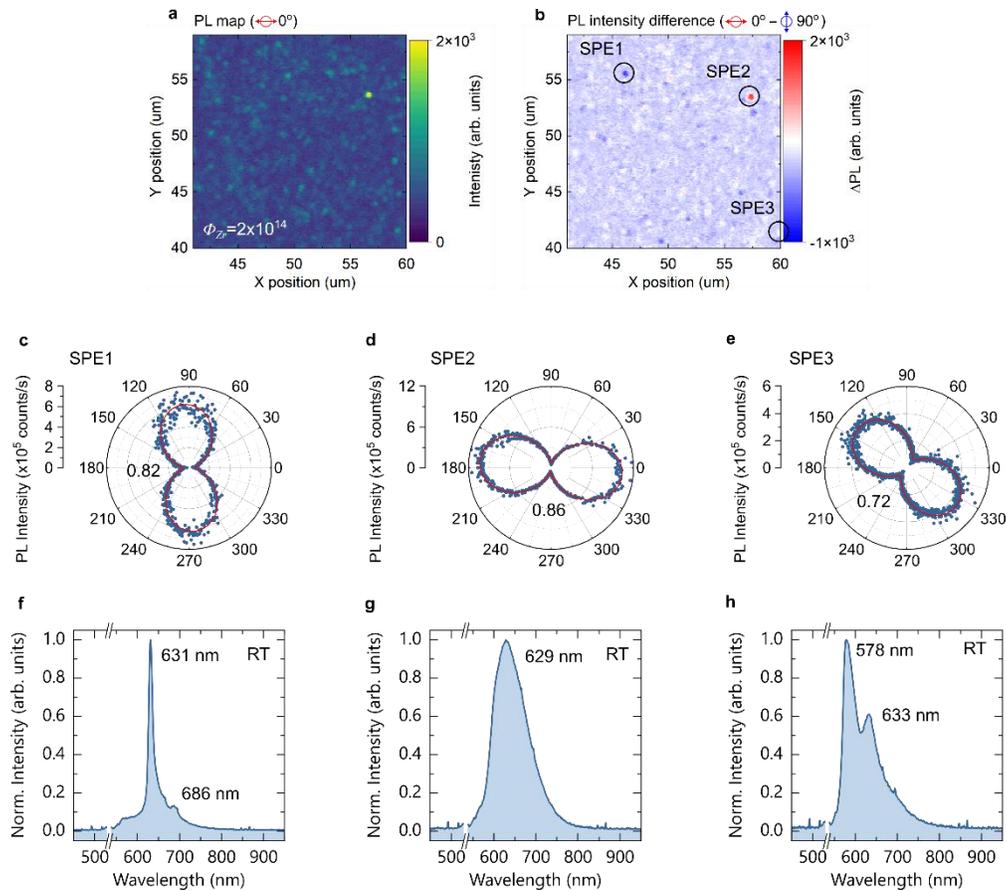

**Figure 5**. Polarization analysis of PL emission. (a) Confocal 20x20 µm² PL intensity map of the Zr-implanted sample D measured at 0° of the polarizer. (b) PL intensity map produced by subtraction of PL intensity maps collected at 90° from the one at 0° of the polarizer rotation. (c-e) Polarization diagrams depicting the PL emission from emitters labeled in (b) with (f-h) corresponding PL spectra.

**Search for Zr-related defects at shorter wavelengths**

All the experiments described above were performed with laser excitation at 532 nm. These measurements did not provide clear evidence of previously anticipated [52] [53] formation of single-photon emitters based on defects incorporating Zr. The current assumption is that with 532 nm excitation we predominantly observe the photoluminescence from single-photon emitters resulting from the damages in the AlN lattice produced by ion implantation. These emitters exhibited spectral characteristics similar to those previously reported in literature attributed to vacancy-based defects within AlN.

In addition to PL spectra from individual single-photon emitters, we measured using the 532nm excitation wavelength the background fluorescence of AlN films away from emitters and found the dependence on implantation fluence (**Fig. S5**, Supplementary Materials). We found the appearance of the strong PL band at about 700 nm with the increase of the implantation fluence (**Fig. S5b**). Notably, this PL peak is more pronounced for Zr-implanted AlN samples compared to the Kr-implanted samples. The rise of an additional PL band and increase of its intensity with ion implantation fluence may indicate the formation of a Zr-based defects in addition to vacancy-type defects due to implantation damage of the AlN lattice.

Interestingly, an additional emission in bulk AlN films implanted with Zr ions at 730 nm (1.7 eV) has been previously observed using a laser source at 266 nm [54]. This emission line was attributed to the formation of Zr-related defect complexes. However, the formation of single-photon emitters was not addressed in that study.

In the work by Varley et al., it was predicted from the first-principles calculations based on density functional theory (DFT) that for the Zr-related defect complex such as $(Zr_{Al}V_N)^0$, the ZPL transition should occur at 2.36 eV or 525 nm [52]. To explore quantum emitters with PL emission at this specific wavelength, we employed a shorter wavelength excitation at 405 nm. We found that the background fluorescence under this excitation wavelength is substantially stronger than with 532 nm laser even for the lowest ion implantation fluence. Nonetheless, we were able to identify isolated bright spots in the sample with the lowest implantation fluence of $\Phi_{Zr} = 5 \times 10^{11}$ (**Fig. 6a**). To mitigate the influence of AlN background fluorescence, we employed both long-pass and short-pass filters. Importantly, the observed isolated emitters in **Fig. 6a** had PL spectra distinct from the AlN background featuring pronounced narrow emission lines at 503 nm and 543 nm, respectively (**Fig. 6b, c**). Such distinctive spectra were absent in the control AlN sample A without implantation. However, it is essential to note that the signal-to-noise ratio presented challenges in measuring second-order autocorrelation histograms necessary to confirm the single-photon nature of these emitters. Consequently, our future investigations will require slightly longer excitation wavelength, closer in proximity to the anticipated zero-phonon-line transition at 525 nm.

Interestingly, the change in excitation wavelength also led to an observable dependence of the AlN background fluorescence on Zr-ion implantation fluence (details provided in Supplementary Materials, **Fig. S6**). However, the most prominent emission peak in background spectra was now centered at approximately 665 nm, which calls for further research beyond this study.

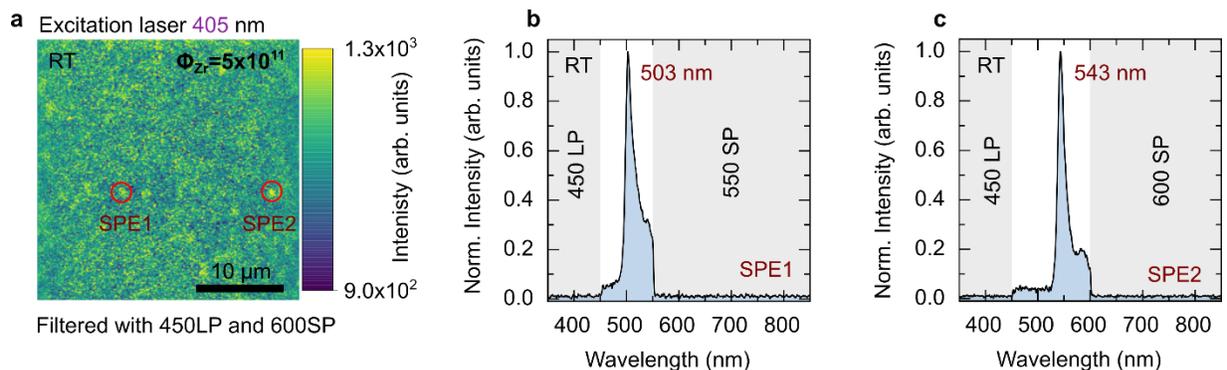

**Figure 6**. Analysis of single-photon emitters at short wavelengths using 405 nm excitation laser at room temperature. (a) PL intensity map of sample B with the lowest implantation fluence $\Phi_{Zr} = 5 \times 10^{11}$. Red circles: isolated point emitters labeled SPE1 and SPE2, respectively. (b) and (c) show PL spectra of SPE1 and SPE2, respectively. The emission is filtered with 450nm long-wavelength pass and 550 nm (600 nm) short-wavelength pass filters and background-corrected.

**Conclusion**

In conclusion, this study provides a comprehensive analysis of single-photon emitters in AlN induced by Zr and Kr heavy ion implantation, followed by thermal annealing. We demonstrate that ion implantation is an effective method for creating single-photon emitters in the AlN lattice for both types of ions. Under the excitation of 532 nm wavelength, the observed single-photon emitters in the implanted samples show similar characteristics with the sparse emitters already present in non-implanted samples. The results of Zr and Kr implantation show similar outcomes, suggesting no incorporation of these ions into the lattice. These emitters are primarily associated with point defects such as vacancy or interstitial defects in the AlN lattice. We observed that the density of emitters increases with ion fluence, with an optimal balance between emitter density and AlN background fluorescence achieved at $\Phi_{Zr} = 2 \times 10^{14}$. However, under shorter wavelength excitation at 405 nm, quantum emitters in Zr-implanted AlN exhibit unique spectral properties, possibly indicating the presence of theoretically predicted optically active Zr-based defect complexes. Future investigations should concentrate on exploring the single-photon emission and spin properties of these defects, particularly focusing on the Zr-based emitters, to unlock their full potential in quantum technologies. This research contributes to our understanding of defect-center related single-photon emitters in AlN and advances the AlN platform's potential for quantum photonic applications.

**Methods**

**Formation of emitters by implantation and thermal annealing.** We used commercially available AlN templates on sapphire ($Al_2O_3$) substrates grown by plasma vapor deposition of nanocolumns (PVDNC$^{TM}$) from Kyma Technologies Inc. The nominal AlN film thickness was 200 nm ± 5%. The 2" wafer was diced into 10 mm by 10 mm samples for subsequent ion implantation.

The implantation was performed on 200 kV Danfysik Research Ion Implanter in Ion Beam Materials Laboratory at Los Alamos National Laboratory (LANL). The implantation energy of 200 keV (190 keV) was selected to restrict the Zr (Kr) implantation depth of ~80 nm following the simulation results obtained with Stopping and Range of Ions in Matter (SRIM) software. We used the ion fluence from $\Phi_{Zr} = 5 \times 10^{11}$ to $\Phi_{Zr} = 2 \times 10^{15}$ to vary the concentration of Zr or Kr ions incorporation. The target temperature was kept at room temperature during all implantations.

The samples were then solvent cleaned and annealed in conventional quartz tube furnace (Blue M Oven Furnace) at 1000C under argon (Ar) atmosphere for 30 minutes. After thermal annealing, the samples were cleaned with solvents and plasma ashing with $O_2$ gas. The formation of single-photon emitters was confirmed then by photoluminescence intensity mapping.

**Experimental setup.** The optical characterization of quantum emitters in AlN samples was conducted at both room temperature and liquid nitrogen (77K) using the freezing microscope stage (THMS600, Linkam). We utilized a custom-built scanning confocal microscope, which was based on a commercially available inverted microscope body (Nikon, Ti-U model). This microscope was equipped with a 100 µm pinhole and a 100x air objective featuring a numerical aperture (NA) of 0.90, all provided by Nikon. For measurements at liquid-nitrogen temperatures, we used a 50x air objective with an NA of 0.60 (Nikon). To facilitate confocal scanning, the objective was mounted on a piezo stage, specifically the Physik Instrumente P-561 model, controlled by a Physik Instrumente E-712 controller and interfaced with LabVIEW from National Instruments. For optical excitation of the emitters, we employed a continuous wave diode-pumped solid-state laser with a power output of 200 mW, emitting at 532 nm (Lambda beam PB 532-200 DPSS, RGB

Photonics). The laser pump spot size on the sample was estimated to be approximately 1 µm. To separate the excitation light from the photoluminescence (PL) signal, we utilized a 550 nm long-pass dichroic mirror (DMLP550, Thorlabs). Any residual pump power was further attenuated using a 550 nm long-pass filter (FEL0550, Thorlabs) positioned in front of the detectors. For short-wavelength excitation, we used a continuous-wave diode laser emitting at 405 nm with an output power of 200 mW (Lambda beam PB 405-200, RGB Photonics). The laser light was filtered from the PL signal using a 409 nm long-pass filter (FF02-409/LP-25, AVR Optics). Emission data was collected using an avalanche detector with a quantum efficiency of 69% at 650 nm (SPCM-AQRH, Excelitas) to enable single-photon detection during the scanning process. To investigate the quantum characteristics of the emitters, we conducted measurements of the second-order autocorrelation function $g^{(2)}(\tau)$ employing a Hanbury Brown and Twiss (HBT) setup. This setup consisted of two avalanche detectors featuring a time resolution of 30 ps and a quantum efficiency of 35% at 650 nm (PDM, Micro-Photon Devices), along with an acquisition card featuring just 4 ps internal jitter (SPC-150, Becker & Hickl).

**Acknowledgements**

This work is supported by the U.S. Department of Energy (DOE), Office of Science through the Quantum Science Center (QSC), and the National Science Foundation Award 2015025-ECCS. Zr and Kr ion implantation is supported by DOE BES grant, LANLE3QR and performed at the at the Center for Integrated Nanotechnologies (CINT), an Office of Science User Facility operated for the U.S. Department of Energy (DOE) Office of Science by Los Alamos National Laboratory. Los Alamos National Laboratory, an affirmative action equal opportunity employer, is managed by Triad National Security, LLC for the U.S. Department of Energy's NNSA, under contract 89233218CNA000001.

**Notes**

While preparing this manuscript, the authors learned about another study that describes single-photon emitters in aluminum nitride, created through the process of aluminum ion implantation, as referenced in [56].

**Supplementary Materials**

**Quantum Emitters in Aluminum Nitride Induced by Zirconium Ion Implantation**

Alexander Senichev[1,2], Zachariah O. Martin[1,2], Yongqiang Wang[2], Owen M. Matthiessen[1], Alexei Lagutchev[1], Han Htoon[2,4], Alexandra Boltasseva[1,2], Vladimir M. Shalaev[1,2*]

[1]Elmore Family School of Electrical and Computer Engineering, Birck Nanotechnology Center and Purdue Quantum Science and Engineering Institute, Purdue University, 610 Purdue Mall, West Lafayette, IN, 47907, USA
[2]Quantum Science Center, Department of Energy, A National Quantum Information Science Research Center of the U.S, Oak Ridge National Laboratory, 1 Bethel Valley Road, Oak Ridge, TN 37830, USA
[3]Materials Science and Technology Division, Los Alamos National Laboratory, Los Alamos, NM, 87545, USA
[4]Materials Physics and Applications Division, Los Alamos National Laboratory, Los Alamos, NM, 87545, USA
*Email: shalaev@purdue.edu

## 1. Photoluminescence intensity maps and representative quantum emitters of Kr-implanted AlN

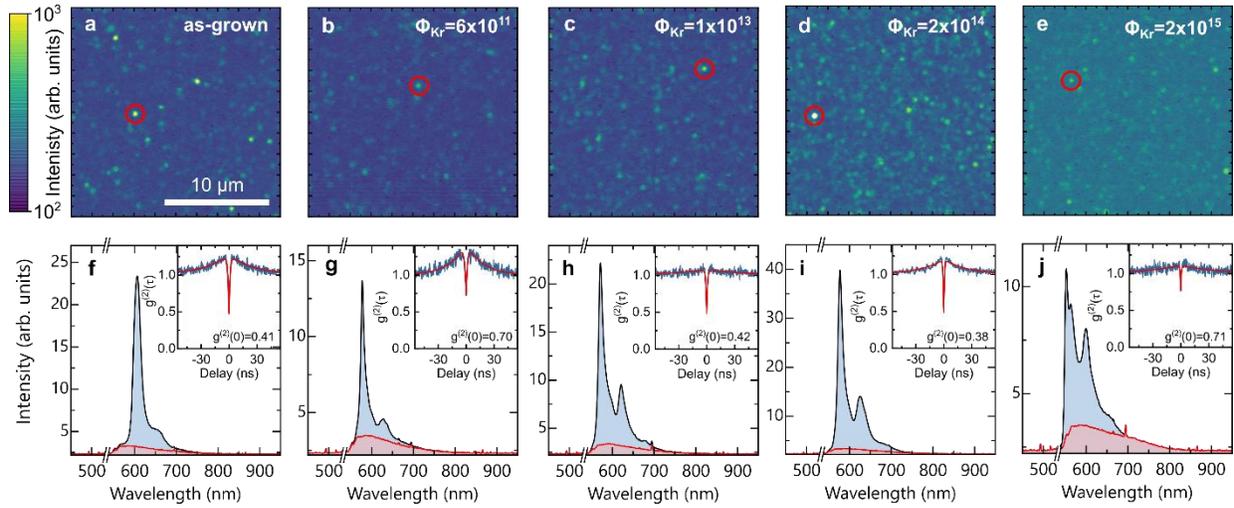

**Figure S1**. PL intensity maps for different Kr-implanted AlN samples with an implantation fluence: (a) as-grown control sample, (b) $\Phi_{Kr} = 6 \times 10^{11}$, (c) $\Phi_{Kr} = 1 \times 10^{13}$ (d) $\Phi_{Kr} = 2 \times 10^{14}$, (e) $\Phi_{Kr} = 2 \times 10^{15}$. (f-j) PL spectra of representative SPEs indicated with red circles in PL intensity maps (a-e). Background PL spectra of AlN for corresponding ion fluences are shown in red. Insets: second-order autocorrelation histograms $g^{(2)}(\tau)$ fitted with a three-level model.

## 2. Examples of room-temperature PL spectra of quantum emitters in Zr-implanted AlN

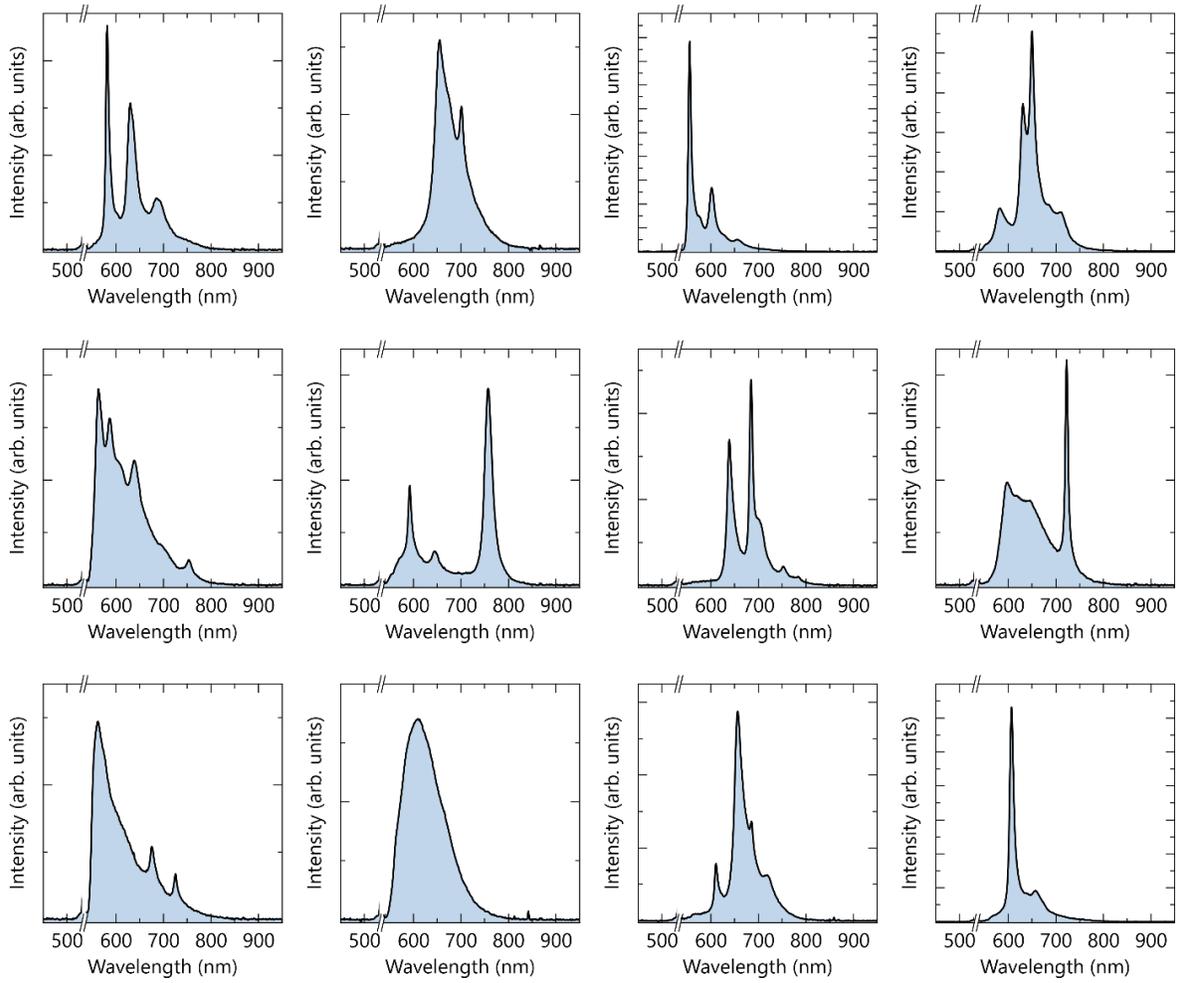

**Figure S2**. Representative PL spectra of different emitters from Zr-implanted AlN at $\Phi_{Zr} = 2 \times 10^{14}$ ions/cm$^2$.

## 3. Wavelength distribution for single-photon emitters in Kr-implanted AlN

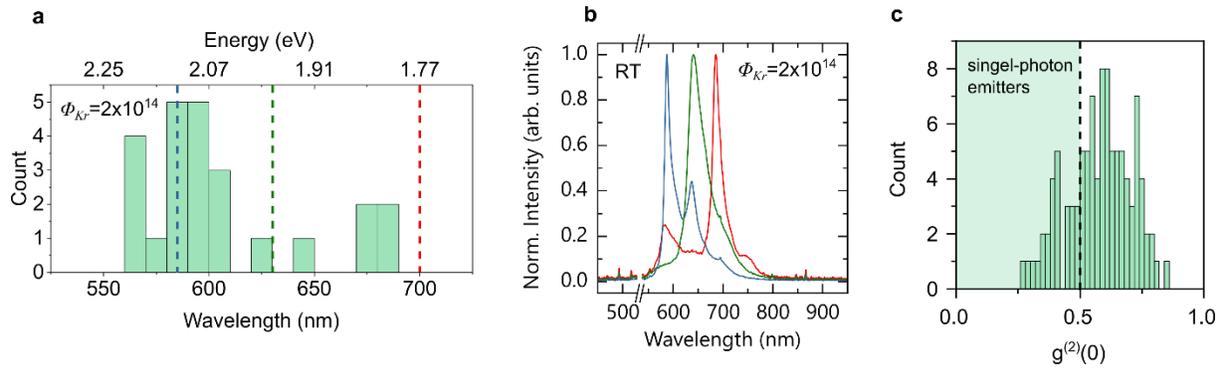

**Figure S3**. (a) ZPL peak wavelength distribution obtained from 24 single-photon emitters in the Kr-implanted (H) sample. Bin size: 10 nm. (b) Representative PL spectra for spectral regions indicated in (a). (c) Histogram of $g^{(2)}(0)$ distribution for emitters in the sample A and sample D. Bin size 0.02. The $g^{(2)}(0)$ values were obtained without background correction and spectral filtering.

## 4. Emission polarization maps of single-photon emitters in AlN

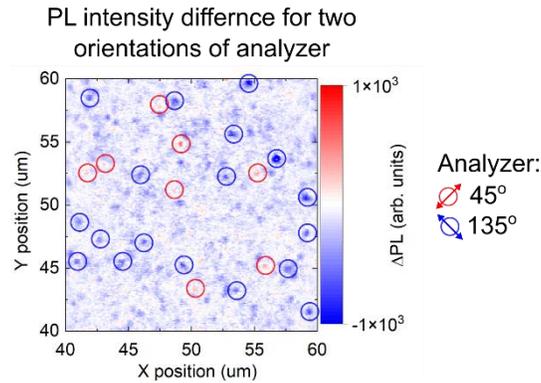

**Figure S4**. Polarization dependent intensity map produced by subtraction of PL intensity map collected at 135° of polarizer orientation from the one at 45°.

## 5. Background data as a control experiment to understand the origin of the defect emission

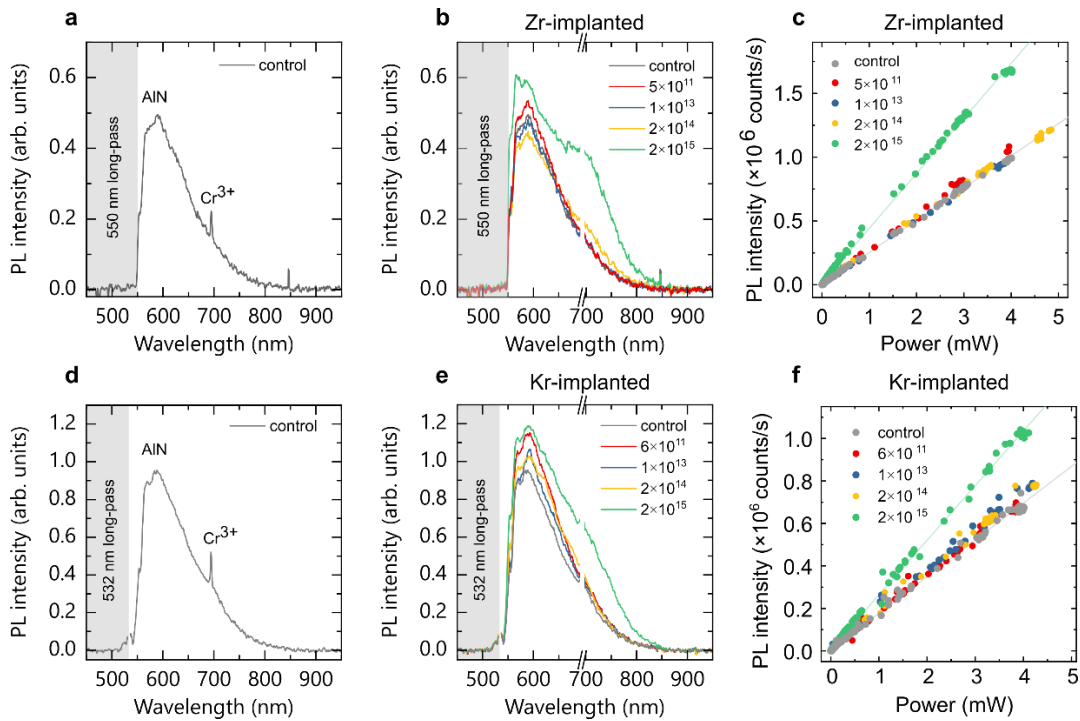

**Figure S5**. Comparison of background (BG) photoluminescence (PL) from annealed AlN films implanted with Zr and Kr ions. (a, d) BG PL spectrum of the AlN film without ion implantation. Sharp peak at 695 nm – chromium ($Cr^{3+}$) impurities in the sapphire substrate [57], (b, e) BG PL spectra of Zr (Kr) implanted AlN films. (c, f) PL intensity as a function of the laser excitation power for different Zr (Kr) implanted AlN films.

To measure the AlN film fluorescence, we collected the PL signal in the areas between bright isolated emitters (e.g., **Fig. 1e,** white circle) and identified the emission from these areas as background fluorescence.

The representative background photoluminescence spectrum of the control as-grown sample (A) is shown in **Fig. S5a**. The PL spectrum was collected above 550 nm using a long-wavelength-pass (LWP) optical filter to suppress the laser line. The PL spectrum shows broad emission band from 550 nm to about 800 nm with a peak around 600 nm. The narrow emission peak at 695 nm is attributed to the chromium ($Cr^{3+}$) impurities in sapphire substrate [57], which is not related to emission mechanisms in AlN film.

The background PL spectra of Zr-ion implanted samples with implantation fluence of $5 \times 10^{11}$ ions/ cm² (B) and $1 \times 10^{13}$ ions/ cm² (C) differ only marginally from the control AlN sample (A) in both PL intensity and the line shape. However, further increase of the ion fluence to $2 \times 10^{14}$ ions/ cm² (D) leads to the rise of PL band around 700 nm. This PL band becomes particularly pronounced at implantation fluence of $2 \times 10^{15}$ ions/cm² (E) (**Fig. S5b**). From the excitation power dependent measurements of the overall background photon count rate, we found that it does not change from the control sample up to an ion fluence of $2 \times 10^{14}$ ions/ cm². The background fluorescence significantly increases for an implantation fluence $2 \times 10^{15}$ ions/ cm² (E) as compared to lower implantation fluences and the control sample (**Fig. S5c**). A similar trend is observed in the background PL spectra of Kr-ion implanted samples. However, the increase in background intensity and the appearance of an additional PL band around 700 nm are less pronounced than in the case of Zr-implanted AlN (**Fig. S5d-f**).

6. **Background fluorescence of Zr-implanted AlN films as a function of ion fluence at 405 nm excitation wavelength**

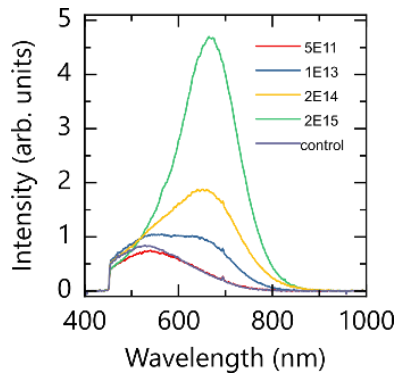

**Figure S6**. Background photoluminescence (PL) under 405 nm excitation wavelength from annealed AlN films implanted with Zr-ions at different implantation fluence. The PL peak intensity at ~665 nm increases with the increase of the implantation fluence.